\documentclass[aps,prl,twocolumn,showpacs,groupedaddress,nofootinbib]{revtex4}  
\usepackage{graphicx}  
\usepackage{dcolumn}   
\usepackage{bm}        
\usepackage{amssymb}   
\usepackage{xspace}
\usepackage{epsfig}

\newcommand{\mpt}{\slash\kern-5pt p_T}

\newcommand{\ttbar}{\mbox{$t\overline{t}$}}
\newcommand{\ppbar}{\mbox{$p\bar{p}$}}
\newcommand{\Ndata}{\mbox{$N_{\rm data}$}}
\newcommand{\Nloose}{\mbox{$N_{\rm loose}$}}
\newcommand{\Ntt}{\mbox{$N_{t\overline{t}}$}}
\newcommand{\NWj}{\mbox{$N_{W{\rm jets}}$}}
\newcommand{\NMC}{\mbox{$N_{\rm other}$}}
\newcommand{\Nlj}{\mbox{$N_{\ell\rm j}$}}
\newcommand{\Njj}{\mbox{$N_{\rm jj}$}}
\begin{document}
\hspace{5.2in} \mbox{FERMILAB-PUB-08-064-E}

\title{Measurement of the \ttbar\ production cross section in \ppbar\ collisions at $\sqrt{s} = 1.96$~TeV}
%
\author{V.M.~Abazov$^{36}$}
\author{B.~Abbott$^{75}$}
\author{M.~Abolins$^{65}$}
\author{B.S.~Acharya$^{29}$}
\author{M.~Adams$^{51}$}
\author{T.~Adams$^{49}$}
\author{E.~Aguilo$^{6}$}
\author{S.H.~Ahn$^{31}$}
\author{M.~Ahsan$^{59}$}
\author{G.D.~Alexeev$^{36}$}
% ...

\affiliation{\vspace{0.1 in}(The D\O\ Collaboration)\vspace{0.1 in}}
% ...
\affiliation{$^{6}$University of Alberta, Edmonton, Alberta, Canada,
                Simon Fraser University, Burnaby, British Columbia, Canada,
                York University, Toronto, Ontario, Canada, and
                McGill University, Montreal, Quebec, Canada}
% ...
\affiliation{$^{29}$Tata Institute of Fundamental Research, Mumbai, India}
% ...
\affiliation{$^{31}$Korea Detector Laboratory, Korea University, Seoul, Korea}
% ...
\affiliation{$^{36}$Joint Institute for Nuclear Research, Dubna, Russia}
% ...
\affiliation{$^{49}$Florida State University, Tallahassee, Florida 32306, USA}
% ...
\affiliation{$^{51}$University of Illinois at Chicago,
                Chicago, Illinois 60607, USA}
% ...
\affiliation{$^{59}$Kansas State University, Manhattan, Kansas 66506, USA}
% ...
\affiliation{$^{65}$Michigan State University,
                East Lansing, Michigan 48824, USA}
% ...
\affiliation{$^{75}$University of Oklahoma, Norman, Oklahoma 73019, USA}
% ...
\date{March 18, 2008}

\begin{abstract}
We measure the \ttbar\ production cross section in \ppbar\ collisions at $\sqrt{s}=1.96$~TeV in the lepton+jets channel. Two complementary methods discriminate between signal and background, $b$-tagging and a kinematic likelihood discriminant. Based on 0.9 fb$^{-1}$ of data collected by the D0 detector at the Fermilab Tevatron Collider, we measure $\sigma_{t\overline{t}}=7.62\pm0.85$~pb, assuming the current world average $m_t=172.6$~GeV. We compare our cross section measurement with theory predictions to determine a value for the top quark mass of $170\pm7$~GeV.
\end{abstract}

\maketitle 

The standard model fixes all properties of the top quark except its mass. The cross section for top quark production depends on the couplings of the top quark and on its mass. In this Letter, we report the most precise measurement of the top-antitop quark pair (\ttbar) production cross section to date. By comparing the measured cross section to predictions we test whether the top quark conforms with standard model expectations. We also for the first time extract a constraint on the top quark mass based only on this comparison. This determination of the top quark mass is complementary to direct measurements and has the advantage that it is done in a well-defined renormalization scheme, that employed in the calculation of the cross section.

The Tevatron collides protons and antiprotons at $\sqrt{s}=1.96$~TeV. Most top quarks at the Tevatron are created in pairs through the strong interaction, although evidence of single top quark production has been reported recently~\cite{single_top}. For a top quark mass of 175~GeV, the standard model predicts a \ttbar\ production cross section of about 6.7 pb~\cite{Cacciari,Kidonakis}. Previous measurements~\cite{D0xsec,CDFxsec} agree with this prediction within their precision of 15\%. Here we present an substantially improved measurement of the \ttbar\ production cross section, based on data collected by the D0 detector~\cite{D0} between August 2002 and December 2005 with an integrated luminosity of 0.9 fb$^{-1}$. 

In the standard model the top quark always decays to a $W$ boson and a $b$ quark. The decay modes of the $W$ boson define the possible final states. Here we focus on the lepton+jets channel in which one $W$ boson decays to $e\nu$, $\mu\nu$, or $\tau\nu$ followed by $\tau\rightarrow e\nu\bar{\nu}$ or $\mu\nu\bar{\nu}$. We refer to such leptons as prompt. The other $W$ boson decays to jets or to $\tau\nu$ followed by a hadronic $\tau$ decay. The branching fraction for this channel is 38\%. 
The D0 detector acquires these events by triggering on an electron or muon and at least one jet with large momentum component transverse to the beam direction ($p_T$). The event selection~\cite{PRD} requires exactly one electron or muon, that is isolated from other objects in the detector, with $p_T>20$~GeV and $|\eta|<1.1$ (for $e$) or $|\eta|<2$ (for $\mu$), missing transverse momentum \mbox{$\mpt >20$~GeV} (for $e$+jets) or 25~GeV (for $\mu$+jets), and at least three jets with $p_T>20$~GeV and $|\eta|<2.5$. The pseudorapidity is defined as $\eta=-\ln[\tan(\theta/2)]$ and $\theta$ is the polar angle with the proton beam. The leading jet must have $p_T>40$~GeV and the lepton $p_T$ and $\mpt$ vectors must be separated in azimuth to reject background events with mismeasured particles. Jets are reconstructed using the Run~II cone algorithm~\cite{cone} with cone size $\sqrt{(\Delta\phi)^2+(\Delta y)^2}=0.5$, in terms of azimuth $\phi$ and rapidity $y$. 
We call this the inclusive lepton+jets sample. Table~\ref{tab:events} gives the number of selected events (\Ndata). The expected $t\overline{t}$ signal accounts only for about 20\% of this sample. Most events originate from other processes that produce prompt leptons and jets (mostly $W$+jets production) and from events with jets which mimic the signature of a lepton. We use two complementary techniques to distinguish the \ttbar\ signal from these backgrounds; $b$-tagging and a kinematic likelihood discriminant.

We model the \ttbar\ signal and all backgrounds with prompt leptons using Monte Carlo (MC) simulations. We carry out the analyses using \ttbar\ events generated at a reference mass of 175~GeV. $W$+jets and $Z$+jets production are generated using the {\sc alpgen}~\cite{ALPGEN} generator and {\sc pythia}~\cite{PYTHIA} for showering. A matching algorithm~\cite{MLM} avoids double counting of final states. Single top production is generated using {\sc singletop}~\cite{SINGLETOP} and {\sc comphep}~\cite{COMPHEP}. Diboson and \ttbar\ production are generated by {\sc pythia}. All simulated events are processed by a detector simulation based on {\sc geant}~\cite{GEANT} and by the same reconstruction programs as the collider data. 

\begin{table}
\begin{tabular}{lr@{}lr@{}lr@{}lr@{}l}
\hline\hline
& \multicolumn{2}{c}{$e$+3 jets} & \multicolumn{2}{c}{$e$+$\geq$4 jets} & \multicolumn{2}{c}{$\mu$+3 jets} & \multicolumn{2}{c}{$\mu$+$\geq$4 jets} \\
\hline
\Ndata  & 1300 & & 320 & & 1120 & & 306 & \\ 
\Nloose & 2592 & & 618 & & 1389 & & 388 & \\ 
\hline
$\epsilon_s$(\%) & 84.8 &$\pm$0.3 & 84.0 &$\pm$1.8 & 87.3 &$\pm$0.5 & 84.5 &$\pm$2.2 \\
$\epsilon_b$(\%) & 19.5 &$\pm$1.7 & 19.5 &$\pm$1.7 & 27.2 &$\pm$5.4 & 27.2 &$\pm$5.4 \\
\hline
\Ntt & 182 &$\pm$20 & 156 &$\pm$17 & 137 &$\pm$15 & 129 &$\pm$14 \\
\NWj & 718 &$\pm$42 &  69 &$\pm$20 & 802 &$\pm$26 & 131 &$\pm$16 \\	
\NMC & 132 &$\pm$15 &  35 &$\pm$4  & 139 &$\pm$15 &  36 &$\pm$4  \\
\Njj & 268 &$\pm$34 &  60 &$\pm$10 &  42 &$\pm$14 &  10 &$\pm$6  \\
\hline\hline
\end{tabular}
\caption{Event counts in the inclusive lepton+jets sample.\label{tab:events}}
\end{table}

We first determine the background from events without prompt leptons in the inclusive lepton+jets sample using loose data samples defined by relaxing the electron identification and the muon isolation requirements. We use simulated events to determine the probability $\epsilon_s$ for leptons from $W$ boson decays that satisfy the loose selection to also pass the selection used for the measurement. We correct this efficiency for known differences between efficiencies observed in the MC simulation and in data. We determine the corresponding efficiency $\epsilon_b$ for misidentified leptons using data selected with the criteria given above except for requiring $\mpt<10$~GeV to minimize contributions from leptons from $W$ boson decays. The number of events in our selected sample is $\Ndata=\Nlj+\Njj$, where \Nlj\ is the number of events with prompt leptons and \Njj\ the number of events without prompt leptons. The number of events in the corresponding loose sample is $\Nloose=\Nlj/\epsilon_s + \Njj/\epsilon_b$. These two equations determine \Njj\ given in Table~\ref{tab:events}. 
We predict the number of events, \NMC, from the smaller background processes (single top, $Z$+jets, and diboson production) using the MC simulation and next-to-leading order cross sections~\cite{bkg_xsec}. 

For the $b$-tag analysis, we start with the expected \ttbar\ cross section to get a first estimate of the number of \ttbar\ events in the sample. After we obtain a cross section as described below we update this estimate using the measured cross section and iterate the cross section calculation until the result is stable. We fix the number of $W$+jets events in the inclusive sample so that the sum of all background and signal contributions equals the observed number of events.

\begin{figure}[t]
\begin{center}
\epsfig{file=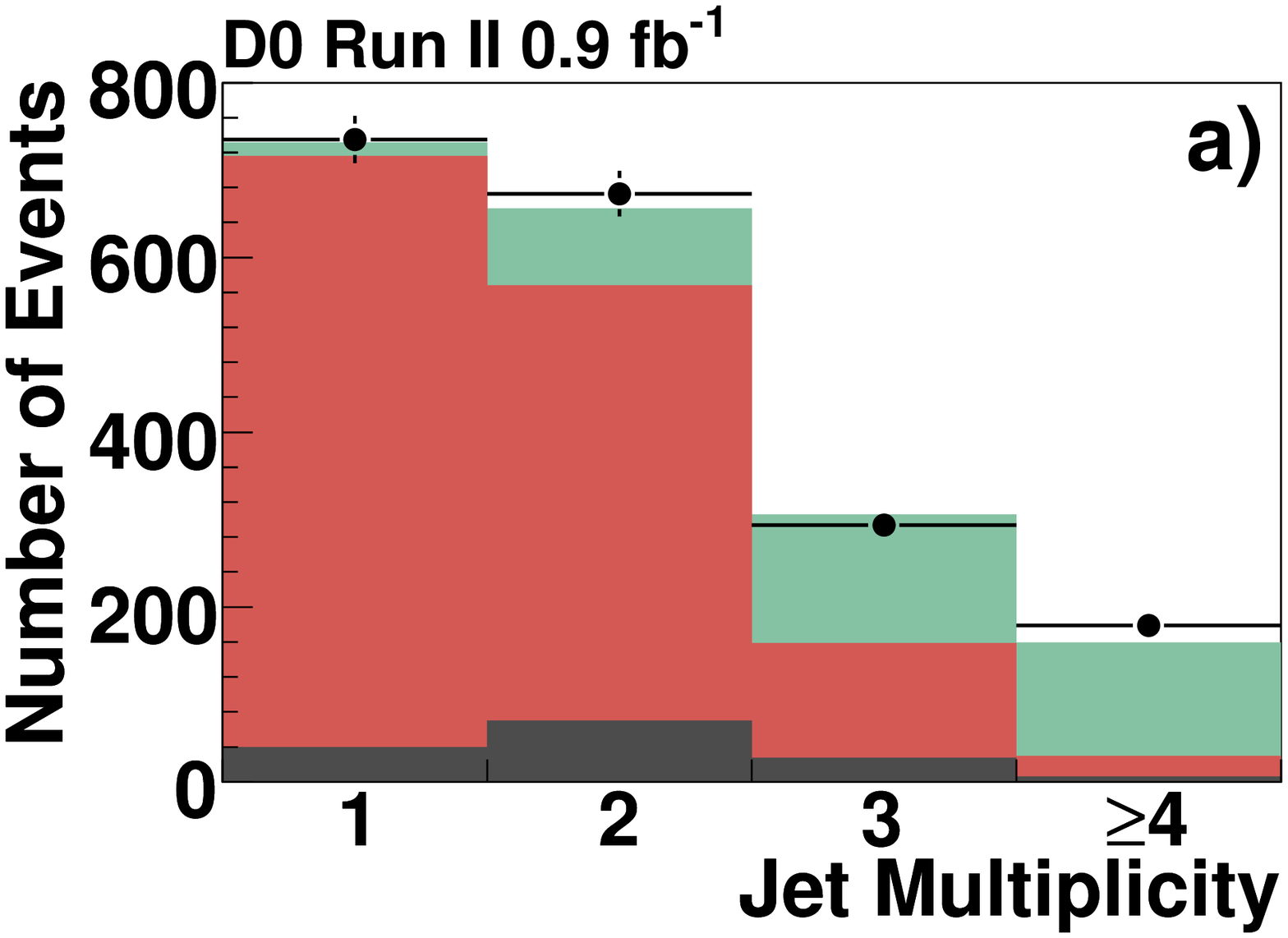,width=1.7in}%
\epsfig{file=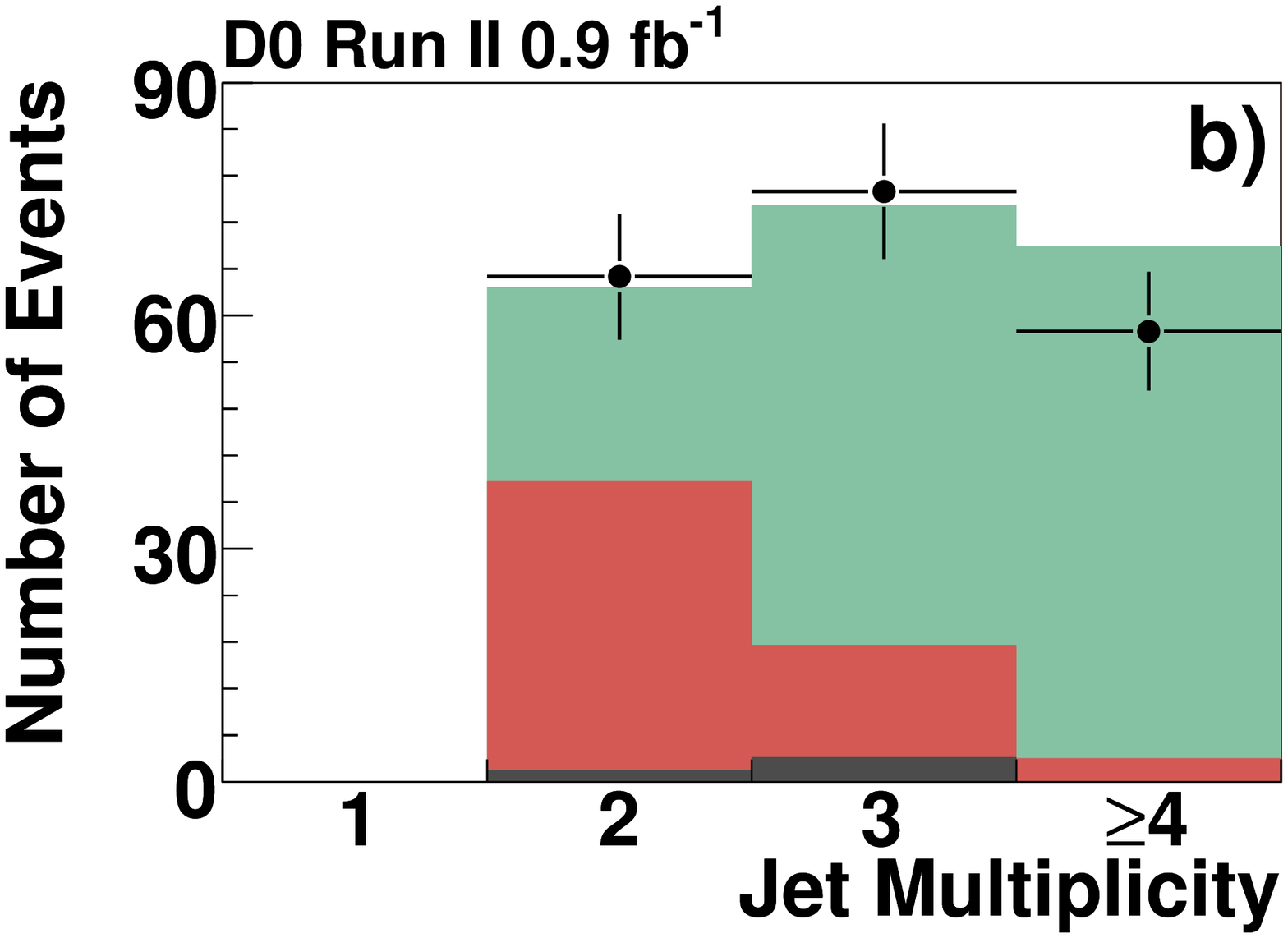,width=1.7in}%
\end{center}
\caption{Jet multiplicity spectra for $e$+jets and $\mu$+jets events (a) with one $b$-tagged jet and (b) with at least 2 $b$-tagged jets. The histogram shows (from top to bottom) the contributions from $t\overline{t}$ production, and from backgrounds with prompt leptons and without prompt leptons.}
\label{fig:Njets}
\end{figure}

The $b$-tag analysis enhances signal purity by requiring that at least one jet be tagged as a $b$-jet, i.e., identified to contain the decay of a long-lived particle such as a $b$-hadron~\cite{btagalgo}. We determine the number of background events without prompt leptons as above and the number of events expected from other background sources from the number of background events in the inclusive sample times their probability to be $b$-tagged. We obtain the $b$-tagging probability from the MC simulation corrected for differences in the $b$-tagging efficiencies observed in the simulation and in data. In order for the MC model to correctly predict the number of lepton+jets events with two jets with at least one $b$-tagged jet, we have to scale the number of $W$+jets events with heavy quarks ($b$, $c$) by a factor of 1.17$\pm$0.18 relative to the rest of the $W$+jets events. We use the same scale factor for $W$+$\geq$3 jet events. Figure~\ref{fig:Njets} shows the jet multiplicity spectrum of events with $b$-tags compared to expectations. The composition of the $b$-tagged samples is given in Table~\ref{tab:btag}. The \ttbar\ contribution in Fig.~\ref{fig:Njets} and Tables~\ref{tab:events} and \ref{tab:btag} is based on the cross section measured in the $b$-tag analysis.

\begin{table}
\begin{tabular}{lr@{}lr@{}lr@{}lr@{}l}
\hline\hline
& \multicolumn{2}{c}{3jets,1tag} & \multicolumn{2}{c}{3jets,$\geq$2tags} & \multicolumn{2}{c}{$\geq$4jets,1tag} & \multicolumn{2}{c}{$\geq$4jets,$\geq$2tags} \\
\hline
\Ntt   &\ \ 147&$\pm$12 &\ \ \ \ 57&$\pm$6 &\ \ \ 130&$\pm$10 &\ \ \ \ \ 66&$\pm$7 \\	
\NWj   & 105&$\pm$5  & 10&$\pm$1 &  16&$\pm$2  &  2&$\pm$1 \\	
\NMC   &  27&$\pm$2  &  5&$\pm$1 &   8&$\pm$1  &  2&$\pm$1 \\	
\Njj   &  27&$\pm$6  &  3&$\pm$2 &   6&$\pm$3  &  0&$\pm$2 \\
\hline
total  & 306&$\pm$14 & 74&$\pm$6 & 159&$\pm$11 & 69&$\pm$7 \\
\Ndata & 294&        & 76&       & 179&        & 58& \\	
\hline\hline
\end{tabular}
\caption{Numbers of events in the $b$-tagged analysis.\label{tab:btag}}
\end{table}

We calculate the cross section using a maximum likelihood fit~\cite{p14} to the number of events in eight different channels defined by lepton flavor ($e$, $\mu$), jet multiplicity (3, $\geq4$), and $b$-tag multiplicity (1, $\geq2$). The likelihood is defined as ${\cal L} = \prod_i {\cal P}(N_i,\mu_i(\sigma_{t\overline{t}}))$, where $i$ runs over the eight channels and ${\cal P}(N,\mu)$ is the Poisson probability to observe $N$ events when $\mu$ are expected. The expected number of events is the sum of the number of events from all backgrounds plus the number of \ttbar\ events as a function of $\sigma_{t\overline{t}}$. We obtain 
$\sigma_{t\overline{t}} = 8.05\pm0.54\mbox{(stat)}\pm0.70\mbox{(syst)}\pm 0.49 \mbox{(lumi)}\mbox{ pb}$ for $m_t=175$~GeV. The third uncertainty arises from the measurement of the integrated luminosity~\cite{lumi}.
 
Table~\ref{tab:syst} lists the systematic uncertainties which arise from the following main categories. {\it Selection} covers acceptance and efficiency for leptons and jets. {\it Jet energy calibration} accounts for jet energy scale and resolution. The $b$-tagging efficiencies for $b$, $c$, and light quark/gluon jets make up the {\it $b$-tagging} uncertainty. {\it MC model} uncertainties originate from the cross sections used to normalize the simulated backgrounds, differences observed between \ttbar\ samples generated with {\sc alpgen} and {\sc pythia}, the factorization and renormalization scale in the $W$+jets simulation, and the parton distributions functions (PDF). \Njj\ covers the determination of the number of events without prompt leptons.

\begin{table}[ht]
\begin{tabular}{lccc}
\hline \hline
source & $b$-tag & likelihood & combined \\
\hline
selection efficiency    & 0.26 pb & 0.25 pb & 0.25 pb\\
jet energy calibration  & 0.30 pb & 0.11 pb & 0.20 pb\\
$b$-tagging             & 0.48 pb & ---     & 0.24 pb\\
MC model                & 0.29 pb & 0.11 pb & 0.19 pb\\
\Njj                    & 0.06 pb & 0.10 pb & 0.07 pb\\
likelihood fit          & ---     & 0.15 pb & 0.08 pb\\
\hline \hline
\end{tabular}
\caption{Breakdown of systematic uncertainties.}
\label{tab:syst}
\end{table}

\begin{table}
\begin{tabular}{ll}
\hline\hline
variable & channel \\ \hline
$\sum_{i=3}^{N_j} p_T(i)$ & all \\ 
$\sum_{i=1}^{N_j} p_T(i)/\sum_{i=1}^{N_j} p_z(i)$ & $e$+3 jets, $e$+$\geq$4 jets\\ 
$\sum_{i=1}^{N_j} p_T(i)+p_T(e)+\mpt$ & $e$+3 jets, $e$+$\geq$4 jets\\
$\Delta R$ between lepton and jet 1 & all\\
$\Delta R$ between jets 1 and 2 & $e$+$\geq$4 jets, $\mu$+$\geq$4 jets\\
$\Delta\phi$ between lepton and $\mpt$ & $\mu$+3 jets, $\mu$+$\geq$4 jets\\
$\Delta\phi$ between jet 1 and $\mpt$ & $e$+3 jets, $\mu$+3 jets\\
sphericity &  all but $\mu$+3 jets\\
aplanarity &  all but $\mu$+3 jets\\
\hline\hline
\end{tabular}
\caption{Variables used for the likelihood discriminant. $\Delta R=\sqrt{(\Delta\phi)^2+(\Delta\eta)^2}$ and $i$ indexes the list of ${N_j}$ jets with $p_T>15$~GeV, ordered in decreasing $p_T$.\label{tab:var}}
\end{table}

The likelihood analysis is based on kinematic differences between events with \ttbar\ decays and backgrounds. No single kinematic quantity can separate signal and background effectively. We therefore build a likelihood discriminant from 5-6 variables, listed in Table~\ref{tab:var}, in each channel. The variables were selected to be well modelled by the MC simulation and to have good discrimination power. For this analysis, we use the inclusive lepton+jets sample with the additional requirement that events with three jets must satisfy $\sum_{i=1}^{N_j} p_T(i)>120$~GeV. The events are divided into four channels defined by lepton flavor and jet multiplicity (3, $\geq$4). 

\begin{figure}[t]
\begin{center}
\epsfig{file=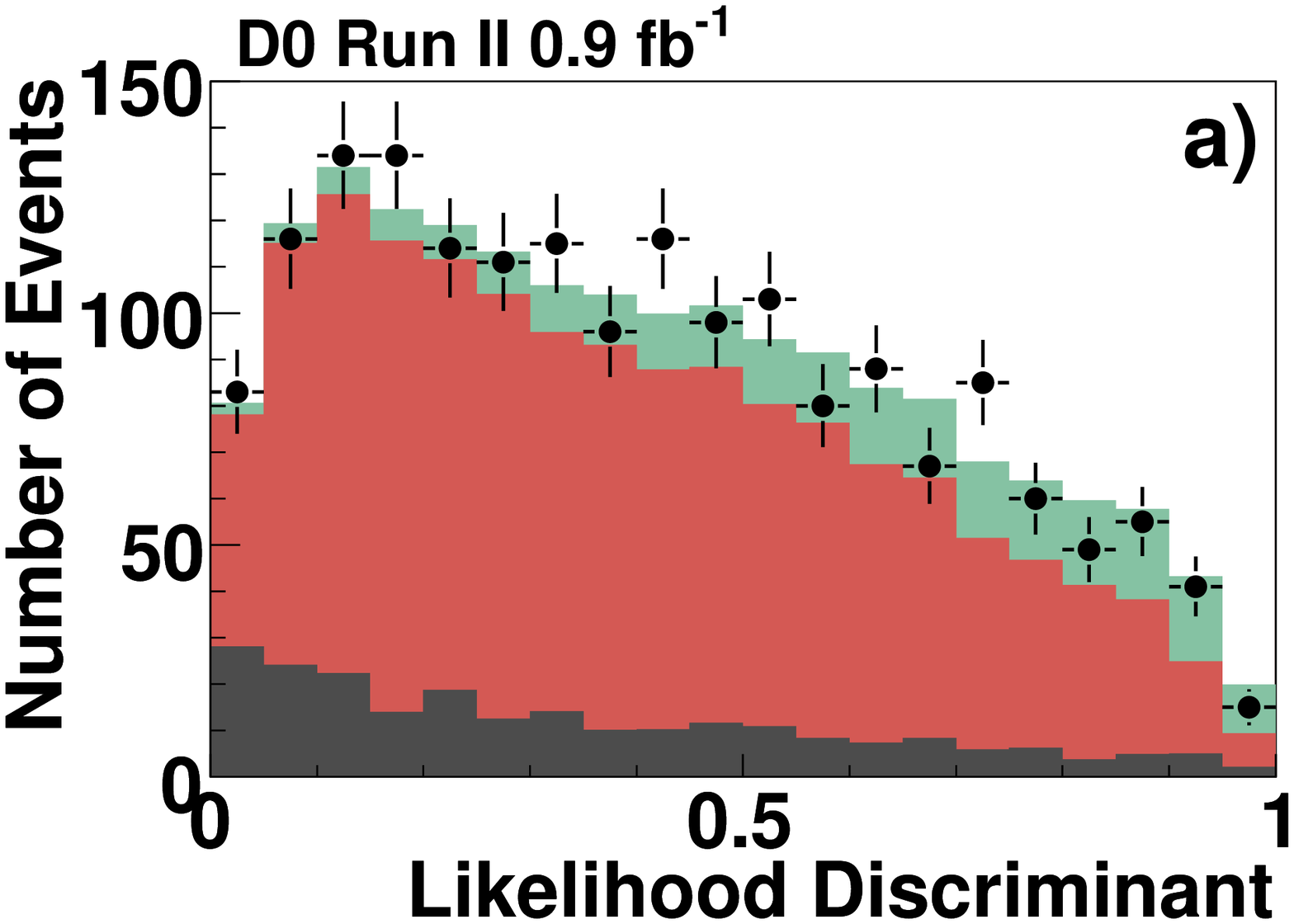,width=1.7in}%
\epsfig{file=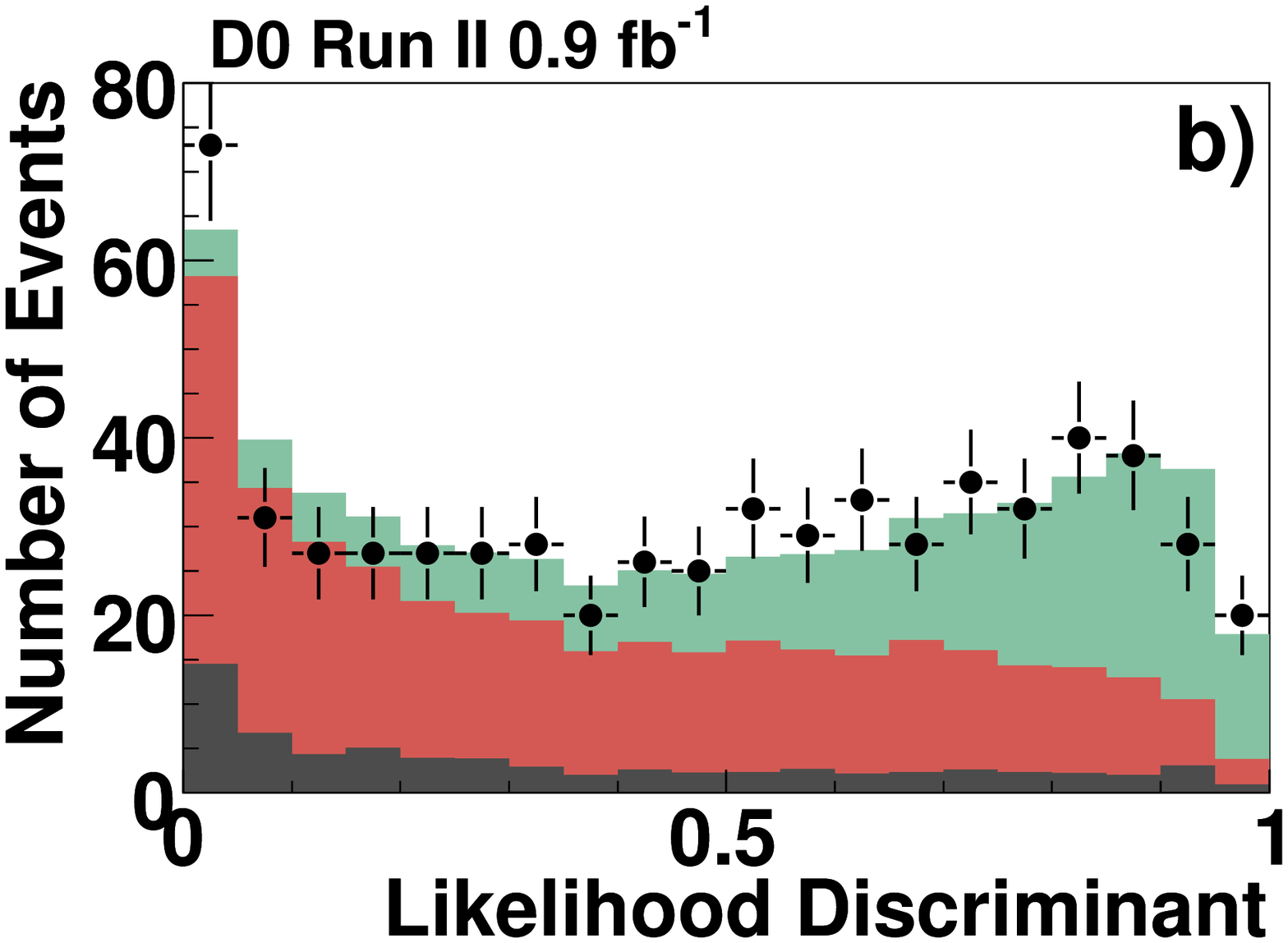,width=1.7in}%
\end{center}
\caption{Likelihood discriminant spectra for $e$+jets and $\mu$+jets events (a) with 3 jets and (b) with at least 4 jets. See also the caption of Fig.~\ref{fig:Njets}.}
\label{fig:like}
\end{figure}

We determine the probability density functions of the likelihood discriminant for signal and prompt lepton backgrounds from the simulation and for events without prompt leptons from a control data sample. We perform a maximum likelihood fit to the likelihood discriminant spectra from data in all four channels simultaneously with the \ttbar\ production cross section as a free parameter. The number of events without prompt leptons is constrained to the value obtained from the loose data sample in the same way as described above. Table~\ref{tab:topo} gives the sample composition for the best fit and Figure~\ref{fig:like} shows the corresponding likelihood discriminant distributions. We measure $\sigma_{t\overline{t}} = 6.62\pm0.78\mbox{(stat)}\pm0.36\mbox{(syst)}\pm 0.40 \mbox{(lumi)}\mbox{ pb}$ for $m_t=175$~GeV. The systematic uncertainties are listed in Table~\ref{tab:syst} in the same categories as for the $b$-tag analysis plus {\it Likelihood fit} which gives the uncertainty from statistical fluctuations in the likelihood discriminant shapes from the MC simulation.

\begin{table}
\begin{tabular}{lr@{}lr@{}l}
\hline\hline
& \multicolumn{2}{c}{3 jets} & \multicolumn{2}{c}{$\geq$4 jets}\\
\hline
\Ndata    &\ 1760&        &\ 626& \\
\Ntt      &  245&$\pm$20\ & 233&$\pm$19\ \\
\NWj+\NMC & 1294&$\pm$48  & 321&$\pm$30\\
\Njj	  &  227&$\pm$28  &  70&$\pm$12\\
\hline\hline
\end{tabular}
\caption{Sample composition from the likelihood fit.\label{tab:topo}}
\end{table}

\begin{figure}[t]
\begin{center}
\epsfig{file=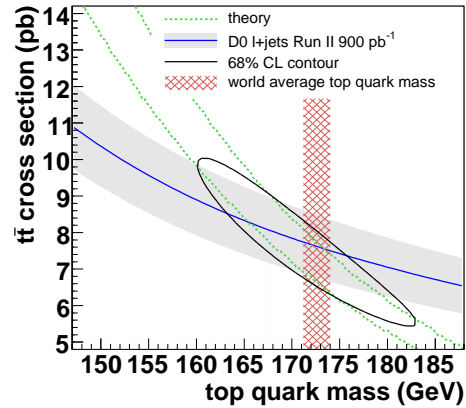,width=2.5in}%
\end{center}
\caption{Comparison of measured cross section and theory prediction versus top quark mass. 
\label{fig:theory}}
\end{figure}

We combine the two analyses using the BLUE method~\cite{BLUE}. Their statistical correlation factor is 0.31, determined by MC generated pseudodata sets that model the statistical correlation between the two analyses. The systematic uncertainties from each source are completely correlated between both analyses. The combined result is $\sigma_{t\overline{t}} = 7.42\pm0.53\mbox{(stat)}\pm0.46\mbox{(syst)}\pm 0.45 \mbox{(lumi)}\mbox{ pb}$ for $m_t=175$~GeV with $\chi^2=2$ for one degree of freedom, corresponding to a p-value of {0.16}. We use samples of \ttbar\ events simulated with different values of the top quark mass to determine the cross section as a function of top quark mass. A polynomial fit gives 
$\sigma_{t\overline{t}}/\mbox{pb}=7.42
-7.9\times10^{-2}\Delta m
+9.7\times10^{-4}(\Delta m)^2
-1.7\times10^{-5}(\Delta m)^3$,
where $\Delta m=m_t/\mbox{GeV}-175$, as shown in Figure~\ref{fig:theory}.
 
We define likelihoods as a function of $\sigma_{t\overline{t}}$ and $m_t$ for the theory prediction and our measurement. There are two sources of uncertainty in the calculated cross sections, a theory uncertainty that arises from the termination of the perturbative calculation and the uncertainty from the PDFs. For each value of $m_t$, we represent the former by a likelihood function that is constant within the ranges given in Refs.~\cite{Cacciari, Kidonakis} and zero elsewhere and the latter by a Gaussian likelihood function with rms equal to the uncertainty determined in Ref.~\cite{Cacciari} for the CTEQ6M~\cite{CTEQ6M} error PDF sets. We then convolute the two functions and average the likelihood functions from the two calculations. 
The cross section measurement is represented by a Gaussian likelihood function centered on the measured value with rms equal to the total experimental uncertainty. We multiply the theory and measurement likelihoods to obtain a joint likelihood. The contour in Figure~\ref{fig:theory} shows the smallest region of the joint likelihood that contains 68\% of its integral. We integrate over the cross section to get a likelihood function that depends only on the top quark mass. We find that at 68\% C.L. $m_t=170\pm7$~GeV, in agreement with the current world average of direct measurements of the top quark mass of $172.6\pm1.4$~GeV~\cite{world_ave}.

In conclusion, we find that \ttbar\ production in $p\overline{p}$ collisions agrees with standard model predictions. At the world average of direct top quark mass measurements of 172.6~GeV we measure $\sigma_{t\overline{t}} = 7.62\pm0.85\mbox{ pb}$. This is the most precise measurement of the \ttbar\ production cross section. By comparing the cross section measurement with the theory prediction we determine the top quark mass to be $170\pm7$~GeV. 
 
%
We thank the staffs at Fermilab and collaborating institutions, 
and acknowledge support from the 
DOE and NSF (USA);
CEA and CNRS/IN2P3 (France);
FASI, Rosatom and RFBR (Russia);
CNPq, FAPERJ, FAPESP and FUNDUNESP (Brazil);
DAE and DST (India);
Colciencias (Colombia);
CONACyT (Mexico);
KRF and KOSEF (Korea);
CONICET and UBACyT (Argentina);
FOM (The Netherlands);
STFC (United Kingdom);
MSMT and GACR (Czech Republic);
CRC Program, CFI, NSERC and WestGrid Project (Canada);
BMBF and DFG (Germany);
SFI (Ireland);
The Swedish Research Council (Sweden);
CAS and CNSF (China);
and the
Alexander von Humboldt Foundation.
%


\begin{thebibliography}{99}
%
\bibitem[a]{alton}
Visitor from Augustana College, Sioux Falls, SD, USA.
\bibitem[b]{burdin}
Visitor from The University of Liverpool, Liverpool, UK.
\bibitem[c]{podesta-lerma}
Visitor from ICN-UNAM, Mexico City, Mexico.
\bibitem[d]{quadt,meyer}
Visitor from II. Physikalisches Institut, Georg-August-University, G{\"o}ttingen, Germany.
\bibitem[e]{voutilainen}
Visitor from Helsinki Institute of Physics, Helsinki, Finland.
\bibitem[f]{wenger}
Visitor from Universit{\"a}t Z{\"u}rich, Z{\"u}rich, Switzerland.

\bibitem[\ddag]{deceased}
Deceased.

%
\vskip 0.25cm

\bibitem{single_top} 
D0 Collaboration, V.M.~Abazov {\sl et al.}, Phys. Rev. Lett. {\bf98}, 
181802 (2007).

\bibitem{Cacciari} 
M. Cacciari {\sl et al.}, JHEP {\bf 404}, 068 (2004).

\bibitem{Kidonakis} 
N. Kidonakis and  R. Vogt, Phys. Rev. D {\bf 68}, 114014 (2003).

\bibitem{D0xsec} 
D0 Collaboration, V.M.~Abazov {\sl et al.}, Phys. Rev. D {\bf 74}, 112004 (2006); Phys. Rev. D {\bf 76}, 052006 (2007).

\bibitem{CDFxsec} 
CDF Collaboration, A.~Abulencia {\sl et al.}, Phys. Rev. Lett. {\bf 97}, 082004 (2006).

\bibitem{D0} 
D0 Collaboration, V.M.~Abazov {\sl et al.}, Nucl. Instrum. Methods Phys. Res., Sect. A {\bf565}, 463 (2006).

\bibitem{PRD} D0 Collaboration, V.M. Abazov {\sl et al.}, Phys. Rev. {\bf D} 76, 092007 (2007).

\bibitem{cone} G.~Blazey {\sl et al.}, arXiv:hep-ex/0005012 (2000). 

\bibitem{ALPGEN} M.L.~Mangano {\sl et al.}, JHEP {\bf 307}, 001 (2003).

\bibitem{PYTHIA} 
T.~Sj{\"o}strand {\sl et al.}, arXiv:hep-ph/0308153 (2003).

\bibitem{MLM}S.~H\"{o}che {\sl et al.}, arXiv:hep-ph/0602031 (2004).

\bibitem{SINGLETOP} E.E.~Boos {\sl et al.}, Phys.\ Atom.\ Nucl.~{\bf 69}, 1317 (2006). 

\bibitem{COMPHEP} E.E.~Boos {\sl et al.} (CompHEP Collaboration), Nucl. Instrum. Methods Phys. Res., Sect. A {\bf534}, 250 (2004). 	 

\bibitem{GEANT}
R.~Brun and F.~Carminati, CERN Program Library Long Writeup W5013 (1993).

\bibitem{bkg_xsec}
E.E.~Boos {\sl et al.}, Phys.\ Atom.\ Nucl.~{\bf 69}, 1317 (2006);
Z.~Sullivan, Phys.\ Rev.\ D {\bf 70}, 114012 (2004);
J.M.~Campbell and R.K.~Ellis, Phys.\ Rev.\ D {\bf 60}, 113006 (1999).

\bibitem{btagalgo}
T.~Scanlon, Ph.D. thesis, FERMILAB-THESIS-2006-43.

\bibitem{p14} D0 Collaboration, V.M. Abazov {\sl et al.}, Phys. Rev. D {\bf 74}, 112004 (2006).

\bibitem{lumi} T.~Andeen {\sl et al.}, FERMILAB-TM-2365 (2007).

\bibitem{BLUE} L.~Lyons, D.~Gibaut and P.~Clifford,  Nucl. Instrum. Methods Phys. Res., Sect. A {\bf 270}, 110 (1988); A.~Valassi,  Nucl. Instrum. Methods Phys. Res., Sect. A {\bf 500}, 391 (2003).

\bibitem{CTEQ6M}D.~Stump {\sl et al.}, J. High Energy Phys. 10 (2003) 046.

\bibitem{world_ave}CDF and D0 Collaborations, FERMILAB-TM-2403-E.

\end{thebibliography}
\end{document}